\documentclass{article}
\usepackage{ijcai22}

\usepackage{times}
\usepackage{soul}
\usepackage{url}

\usepackage[utf8]{inputenc}
\usepackage[small]{caption}
\usepackage{graphicx}
\usepackage{amsmath}
\usepackage{amsthm}
\usepackage{amssymb,amsfonts}
\usepackage{booktabs}
\usepackage{algorithm}
\usepackage{algorithmic}

\usepackage{multirow}
\usepackage{threeparttable}
\usepackage{tabularx} 
\usepackage{colortbl}  
\usepackage{bm}
\usepackage[hidelinks]{hyperref}
\usepackage{url}

\usepackage{pifont}

\title{Unsupervised Voice-Face Representation Learning by \\ Cross-Modal Prototype Contrast}

\author{
Boqing Zhu\and
Kele Xu\footnote{Corresponding author}\and
Changjian Wang\and
Zheng Qin\and 
Tao Sun\and\\
Huaimin Wang\And
Yuxing Peng
\affiliations
College of Computer, \\
National University of Defense Technology, Changsha, China\\
\emails
\{zhuboq, kelele.xu\}@gmail.com
}

\begin{document}

\maketitle

\begin{abstract}
We present an approach to learn voice-face representations from the talking face videos, without any identity labels. Previous works employ cross-modal instance discrimination tasks to establish the correlation of voice and face. These methods neglect the semantic content of different videos, introducing false-negative pairs as training noise.
Furthermore, the positive pairs are constructed based on the natural correlation between audio clips and visual frames. However, this correlation might be weak or inaccurate in a large amount of real-world data, which leads to deviating positives into the contrastive paradigm.
To address these issues, we propose the cross-modal prototype contrastive learning (CMPC), which takes advantage of contrastive methods and resists adverse effects of false negatives and deviate positives. On one hand, CMPC could learn the intra-class invariance by constructing semantic-wise positives via unsupervised clustering in different modalities. On the other hand, by comparing the similarities of cross-modal instances from that of cross-modal prototypes, we dynamically recalibrate the unlearnable instances' contribution to overall loss. Experiments show that the proposed approach outperforms state-of-the-art unsupervised methods on various voice-face association evaluation protocols. Additionally, in the low-shot supervision setting, our method also has a significant improvement compared to previous instance-wise contrastive learning.

\end{abstract}


\section{Introduction}
\label{sec:intro}

\begin{figure}[t]
\centering
\includegraphics[width=0.9\columnwidth]{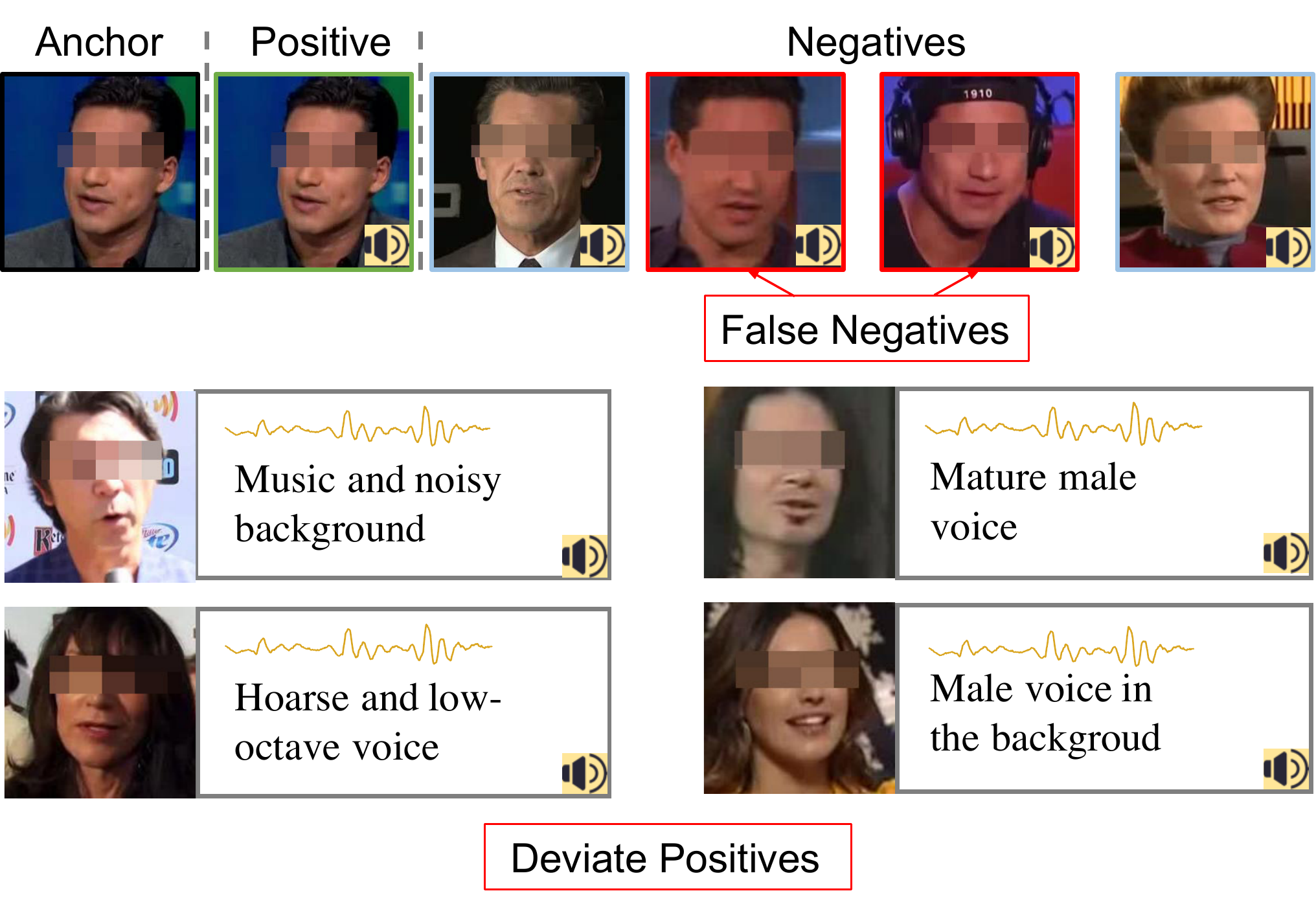}
\caption{False Negatives and Deviate Positives examples in Cross-modal Instance Discrimination. (1) Without identity labels, negative pairs from different videos could actually belong to the same identity, such as the anchor face and negative voice in red border, which leads to false negatives. 
(2) The voice-face correlation might be personalized or irrelevant caused by the noise in the videos, which leads to deviate positives.
The voice modality is displayed by marking a trumpet symbol in the image. 
}
\label{fig:teaser}
\end{figure}

\begin{figure*}[t]
\centering
\includegraphics[width=1.9\columnwidth]{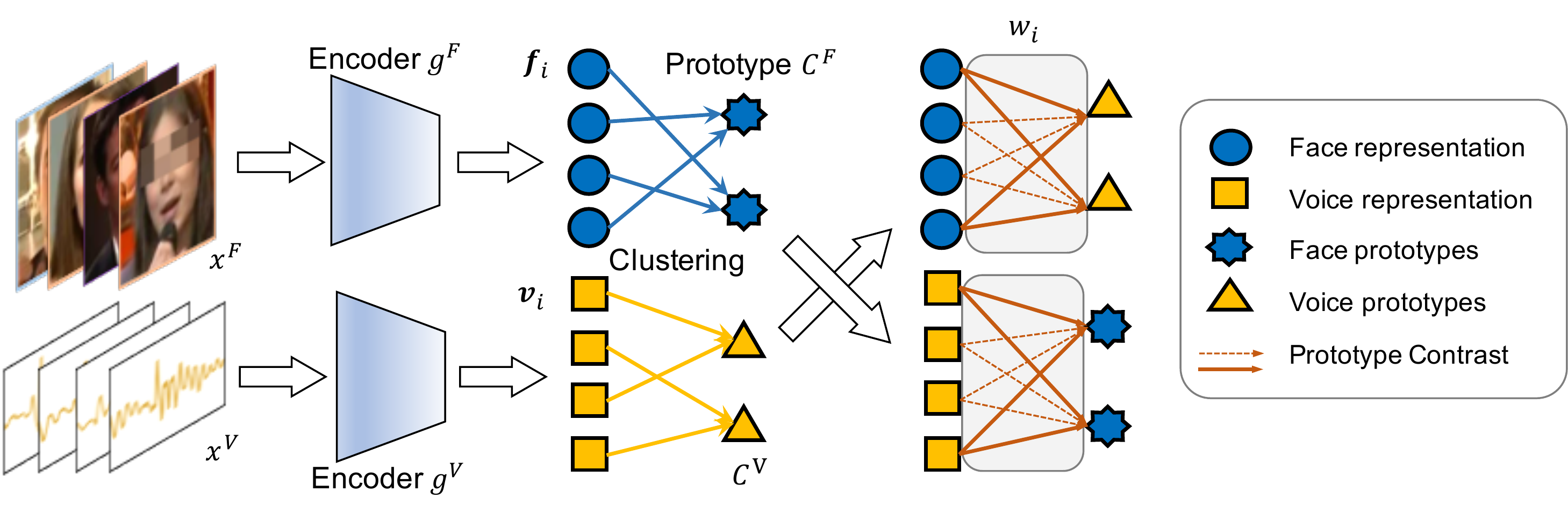}
\caption{\textbf{Overview of the proposed method}. Face frames and voice clips are encoded into corresponding representations. To address the false negatives, clustering is conducted separately on different modalities to learn the prototypes of each instance, aiming to construct semantic-wise positives. Deviate positives are discounted by down-weighting instances.}
\label{fig:pipeline}
\end{figure*}

Neuroimage studies \cite{joassin2011cross} suggest that the interactions between voice and face are not only at a perceptual level, but also in the person recognition process. 
Intuitively, when hearing a person's voice, even if we don't see the person, we immediately have its appearance in our minds. Even for a stranger we have never seen before, we could `construct' what he looks like through his voice. 
While, whether voice and face are relevant and whether they can be perceived by the computational models? 
In recent years, researchers made a sustainable effort on an enduing neural network the ability to recognize faces with the same identity only by hearing the voice or recognizing the voice from the face. It could be formulated as a voice-face association problem. The problem is an interesting scientific study on its own, and it could also benefit practical applications such as virtual reality, criminal investigation, multimodal information retrieval, \textit{etc}.

Psychological and behavioral studies show that voice and face integration skills could emerge very early in our life by processing a number of auditory-visual examples \cite{bahrick2005development}, without intended discipline. This indicates that our computational models may be feasible to learn the association between voice and face in an unsupervised manner.
More recently, contrastive learning has become a prominent tool in unsupervised representation learning, leading to state-of-the-art results in many fields such as vision, language, and speech. A rock line of papers \cite{morgado2021agreement,zolfaghari2021crossclr,morgado2021robust,alayrac2020self} have employed the \textit{Cross-modal Instances Discrimination} (CID) method to learn the multi-modal representations.

In the CID methods, the representations are learned by two opposite forces, the repellence of negative pairs and the attraction of positive pairs. The positive pairs are generated by the audio and images extracted from the same video, and the negative pairs are those extracted from different videos. Then, a contrastive loss (e.g. InfoNCE \cite{oord2018CPC}) is used to pull close the positive pairs and push apart the negative pairs. Despite their improving performance, the CID methods are still confronted with two challenges:
(1) False negative problem. The negative pairs are sampled from different videos regardless of their semantics. For example, two videos have the same \textit{identity} (person), the representations of the voice and face should be pulled close, but they have undesirably pushed apart because they are sampled as negative pairs, as in Figure \ref{fig:teaser}. Since the current CID methods need to utilize a large number of negatives, the problem of the false negative is non-ignorable.
(2) Deviate positive problem. Previous methods have learned the representations by encouraging the network to memorize all the voice-face associations in the dataset. While they ignore the biometric specificity across different identities. Some identities hold an `untypical' or personal voice face correspondence. Moreover, a large amount of video data inevitably contain noise, including blocked faces, excessive head rotation, background music, the voice from outside of the screen, \textit{etc}. We named them `deviate positives' instead of `false positives', because they are all reasonable existing, and to distinguish from supervised learning where there are ground-truth labels.

We firstly reveal the impact of the false negatives and deviate positives in Section \ref{sub:preliminary} from a theoretical view. To combat the above issues, we then propose a Cross-modal Prototype Contrastive Learning (CMPC) algorithm, which takes advantage of previous contrastive methods meanwhile resisting adverse effects of false negatives and deviate positives. Specifically, we obtain the semantic structure by clustering on each modality, and each cluster centroid is used as the prototype of the corresponding instance. The learned prototypes are used to construct the semantic positives with the instance in another modality. 
We use an online clustering approach, which keeps the consistency of instance representations and reduces computation. Additionally, we dynamically down-weight the deviate positives' contribution to overall loss by an Instance Recalibrate method. By combining our methods, we can outperform previous state-of-the-art unsupervised methods on various voice-face association evaluation protocols. Moreover, with low-shot supervision, our method achieves significant improvement compared to previous instance-wise contrastive learning, even outperforms fully supervised models.

The rest of this paper is organized as follows. We first introduce the related work in the following Section \ref{sec:related_work} , then we analyze the problems and present the details of the proposed approach in Section \ref{sec:method}. The experiments are given in the Section \ref{sec:experiments}. Finally, we conclude our paper in Section \ref{sec:conclusion}.


\section{Related Work}
\label{sec:related_work}

\subsection{Learning Association of Voice and Face}
In recent years, the interests for the human voice-face cross-modal learning tasks have surged, and the attempts have shown the feasibility of recognizing a person's face from his voice or identifying a speaker only by giving his face image.
SVHF \cite{nagrani2018svhf} firstly raised voice-face matching and retrieval task. They concatenate the feature from different modalities and then learn a $N$-way classification task by minimizing a cross-entropy loss on the personal identity label.
Instead of explicitly learning the joint embedding between voice and face, DIMNet \cite{wen2018dimnet} learn a shared representation under supervision from one or more covariates, such as gender, ethnicity, and age, by a multi-task object function.
\cite{horiguchi2018face} and \cite{kim2018learning} use the $N$-pair loss and triplet loss to increase the voice-face similarity of the same person respectively.
In \cite{chung2020facefilter}, they focus on the audio-visual speech separation task by using the still face images. \cite{nawaz2021cross} tries to explore the languge dependent problem in voice-face association. For example, the answers question about: “Can a speaker be recognized irrespective of the spoken language?” 
In \cite{wen2021seeking}, they used a two-level inter-modality and intra-modality alignment. Also, they propose a reweighting scheme to better handle the hard samples.
In \cite{nagrani2018learnable}, they learn the cross-modal representation in a self-supervised manner by the contrastive loss, and a curriculum learning schedule boosts the performance.

\subsection{Contrastive Learning}
Contrastive learning has recently become a prominent tool in unsupervised representation learning, leading to state-of-the-art results in many fields such as vision, language and audio.
Contrastive learning either aims to predict parts of the data, or to discriminate between individual instances. 
The former explicitly define a proxy tasks, such as colorization, jigsaw solving, temporal coherence and future prediction \cite{han2020memory,pan2021videomoco}. The later, instances discrimination can be formed as a $N$-classification task which $N$ is the number of instances.
SupCon \cite{khosla2020supcon} extend the self-supervised contrastive approach to the fully-supervised setting, allowing us to effectively leverage label information. Compared to standard supervised classification loss, SupCon shows the robustness to the natural corruption.
The main difference between different contrastive learning approaches lies in their strategy for definition of positive pairs. For example, image random cropping and rotation, or different augmentation of the same instance. 
After getting the positive pairs, comparing with as many negative examples is important to boost the performance. Other works focus on the hard negatives mining \cite{Kalantidis2020mixing} or unbiased negatives sampling \cite{Chuang2020Debiased,wu2021conditional} in contrastive learning.

Clustering-based contrastive learning \cite{caron2018deepclustering,caron2020SwAV,zhan2020online} has been proposed to learn between instance embeddings and cluster assignments.
Among them, PCL \cite{li2021prototypical} formulates the clustering and representation learning as an Expectation-Maximization (EM) algorithm, which approximates and maximizes the log-likelihood function.
While, whether we can learn effective representations if prototypes and instance vectors are derived from different modalities will be explored in this paper.
Closer to our work, XDC \cite{alwassel2020self} performs iterative clustering and classification to learn cross-modal representation. The cluster assignments of one modality are considered as pseudo-label of another modality, following a classification objective. While, the cross-entropy classification loss is sensitive to the noisy label, the inaccurate pseudo-label make it difficult for convergence.
However, our method is conceptually different from the XDC. In XDC, the cluster assignments of one modality are considered as pseudo-label of another modality, following a classification objective, which results in two weaknesses: (1) The cross-entropy classification loss is sensitive to the noisy label, the inaccurate pseudo-label make it difficult for convergence. (2) The initialization of pseudo-label gives endow the semantic information to the cluster centroid, which may lead to training bias. In our method, we actively avoid the pseudo-label and noisy label problem through the clustering and contrast iteration.


\section{Methodology}
\label{sec:method}

\subsection{Preliminary}
\label{sub:preliminary}

Consider a dataset of $N$ video instances $\mathcal{X} = \{(x_i^V, x_i^F) \mid i=1, ..., N\}$ where each instance have two modalities, i.e., voice and face.
The goal of voice-face association is to jointly learn high-level representations $(\textbf{v}_i, \textbf{f}_i)$ for both modalities to capture cross-modal correlations.
The voice and the face of the same \emph{identity} (person) are expected to have similar representations and vice versa.
To this end, let $g^M: x^M \rightarrow \mathcal{R}^{d_m}$ be a parametrized modality specific backbone neural network that takes $x^M$ as input from modality $M \in \{V, F\}$ and outputs a representation of $d_M$ dimension. All the representations are normalized onto a unit hypersphere.
In this work, we follow the unsupervised setting in \cite{nagrani2018learnable} to train the encoders using only video instances without the exact identity labels.


State-of-the-art unsupervised audio-visual representation learning methods \cite{alayrac2020self,morgado2021agreement,patrick2020multi} usually follow the paradigm of \textit{Cross-modal Instance Discrimination} (CID).
For a batch of different video instances, CID uses the voice-face pairs from the same instance as the positive pairs and those from different instances as the negative pairs.
The representations are then trained in a contrastive learning manner \cite{he2020momentum,chen2020simple} across the two modalities.
For simplicity, we only describe the case from voice to face and the same goes for the other side.
Given a positive pair $(\textbf{v}_i, \textbf{f}_i)$, CID establishes a negative face representation set $\mathbb{N}^F_i = \{\textbf{f}_j \mid j \neq i\}$ and seeks to maximize the mutual agreement between $\textbf{v}_i$ and $\textbf{f}_i$ by minimizing the InfoNCE loss \cite{oord2018CPC}:
\begin{equation}
\label{eq:InfoNCE}
\ell_c(\mathbf{v}_i, \mathbf{f}_i) = -\log \frac{\exp(\mathbf{v}_i^T \mathbf{f}_i /\tau)}{\exp(\mathbf{v}_i^T \mathbf{f}_i /\tau) + \sum_{\mathbf{f}_j \in \mathbb{N}^F_i}\exp(\mathbf{v}_i^T \mathbf{f}_j / \tau)},
\end{equation}
where $\tau$ is the temperature hyper-parameter. This process can be interpreted as maximizing the similarity $s^p_i = \mathbf{v}_i^T \mathbf{f}_i$ between the positive pair while suppressing the similarity $s^n_{i,j} = \mathbf{v}_i^T \mathbf{f}_j$ between the negative pairs.
Moreover, the gradients of $s^p_i$ and $s^n_{i,k}$ are computed as:
\begin{align}
\label{eq:sn-gradient}
\frac{\partial \ell_c}{\partial s^n_{i,k}}  & = \frac{1}{\tau}\frac{\exp(s^n_{i,k} / \tau)}{\exp(s^p_i / \tau) + \sum\nolimits_{\mathbf{f}_j \in \mathbb{N}^F_i}\exp(s^n_{i, j} / \tau)}, \\
\label{eq:sp-gradient}
\frac{\partial \ell_c}{\partial s^p_i} & = -\frac{1}{\tau}\frac{\sum\nolimits_{\mathbf{f}_j \in \mathbb{N}^F_i}\exp(s^n_{i, j} / \tau)}{\exp(s^p_i / \tau) + \sum\nolimits_{\mathbf{f}_j \in \mathbb{N}^F_i}\exp(s^n_{i, j} / \tau)}.
\end{align}

As mentioned in Section \ref{sec:intro}, CID suffers from the false negative and the deviate positive problems.
On one hand, for a false negative $(\textbf{v}_i, \textbf{f}_k)$, their similarity $s^n_{i,k}$ could be high as they share the same identity.
From Eq.~\ref{eq:sn-gradient}, this will impose a heavy penalty which pushes $\textbf{v}_i$ away from $\textbf{f}_k$. This is suboptimal as $\textbf{v}_i$ and $\textbf{f}_k$ should be matched together.
On the other hand, for a deviate positive, the positive similarity $s^p_i$ tends to be small. According to Eq.~\ref{eq:sp-gradient}, there will be a heavy penalty which pulls the deviate positive close, which could lead to local optima and hamper the generalizability of the representations.

\subsection{Cross-modal Prototype Contrast}
\label{sub:cmpc}

To tackle false negative problem, we propose a cross-modal prototype-based contrastive learning method.
Our motivation is to extract a set of prototypes which encodes the underlying semantic structure of the whole dataset by clustering the video instances and contrast the instances with the prototypes.
For each instance, its representation in one modality is pulled close to the prototype where it belongs in the other modality and pushed away from other prototypes.
As false negatives usually belong to the same prototype, they will be optimized to have similar representations.

Following \cite{li2021prototypical}, we conduct clustering and contrast iteratively: the prototypes are reconstructed after each training epoch and used in the next epoch.
However, as the encoders change rapidly during training, directly gathering the representations from different training batches induces severe representation inconsistency, which harms the quality of the prototypes.
For this reason, we design \emph{instance feature memories} $(\mathbf{Mem}^V_i, \mathbf{Mem}^F_i)$ for each video instance $x_i$.
The instance feature memories are defined as the momentum update of the instance representations in both modalities:
\begin{align}
\label{eq:memory}
\mathbf{Mem}_i^V & = m \cdot \mathbf{Mem}_i^V+(1 - m)\cdot \mathbf{v}_i, \\
\mathbf{Mem}_i^F & = m \cdot \mathbf{Mem}_i^F+(1 - m)\cdot \mathbf{f}_i,
\end{align}
where $m \in [0, 1)$ is a momentum coefficient.
Compared with $(\mathbf{v}_i, \mathbf{f}_i)$, the change of $(\mathbf{Mem}^V_i, \mathbf{Mem}^F_i)$ are much smoother during training, which provides better consistency.
After each training epoch, we cluster all instances into $K$ distinct groups in each modality based on their instance feature memories.
Here we simply adopt the $k$-means algorithm but we note that our method is agnostic to clustering methods.
The clustering centroids $\mathcal{C}^V = \{\mathbf{c}^V_i \mid i=1,...,K\}$ and $\mathcal{C}^F = \{\mathbf{c}^F_i \mid i=1,...,K\}$ are used as the prototypes.

Next, we apply the contrastive loss between the instance representations and the semantic prototypes in a cross-modal way.
For each instance representation $(\mathbf{v}_i, \mathbf{f}_i)$, we pull $\mathbf{v}_i$ close to the face prototypes $\mathbf{c}^F_s$ where $\mathbf{f}_i$ belongs and push it away from all other face prototypes:
\begin{equation}
\ell_p(\mathbf{v}_i, \mathcal{C}^F) = -\log \frac{\exp(\mathbf{v}_i^T \mathbf{c}_s^F /\tau)}{\exp(\mathbf{v}_i^T \mathbf{c}^F_s / \tau) + \sum_{j\neq s}{\exp(\mathbf{v}_i^T \mathbf{c}_j^F / \tau)}}.
\end{equation}
The same computation goes for $\ell_p(\mathbf{f}_i, \mathcal{C}^V$).
To ensure robust semantic prototype estimation, we cluster the instance feature memories for $R$ times with different numbers of clusters and aggregate the instance-wise contrastive loss to retain local smoothness.
The overall CMPC loss is defined as:
\begin{align}
\mathcal{L}_{\mathrm{CMPC}} = \hspace{3pt} & \ell_c(\mathbf{v}_i, \mathbf{f}_i) + \frac{1}{R}\sum\nolimits_{r=1}^R\ell_p(\mathbf{v}_i, \mathcal{C}_{r}^{F}) \notag \\
& + \ell_c(\mathbf{f}_i, \mathbf{v}_i) + \frac{1}{R}\sum\nolimits_{r=1}^R\ell_p(\mathbf{f}_i, \mathcal{C}_{r}^{V}).
\end{align}

Compared to previous clustering-based contrastive learning methods \cite{caron2018deepclustering,li2021prototypical}, our CMPC differs in the following two aspects:
(1) They adopt a offline update strategy where the representations of all training instances are recomputed before clustering to achieve representation consistency.
This induces heavy computation cost, especially in large-scale datasets.
On the contrary, our method leverages an online strategy which achieves representation consistency without apparent computation overhead.
(2) Our method is designed for cross-modal contrastive learning, while these methods mainly focus on one modality.

\begin{figure}[tp]
\centering
\includegraphics[width=0.9\columnwidth]{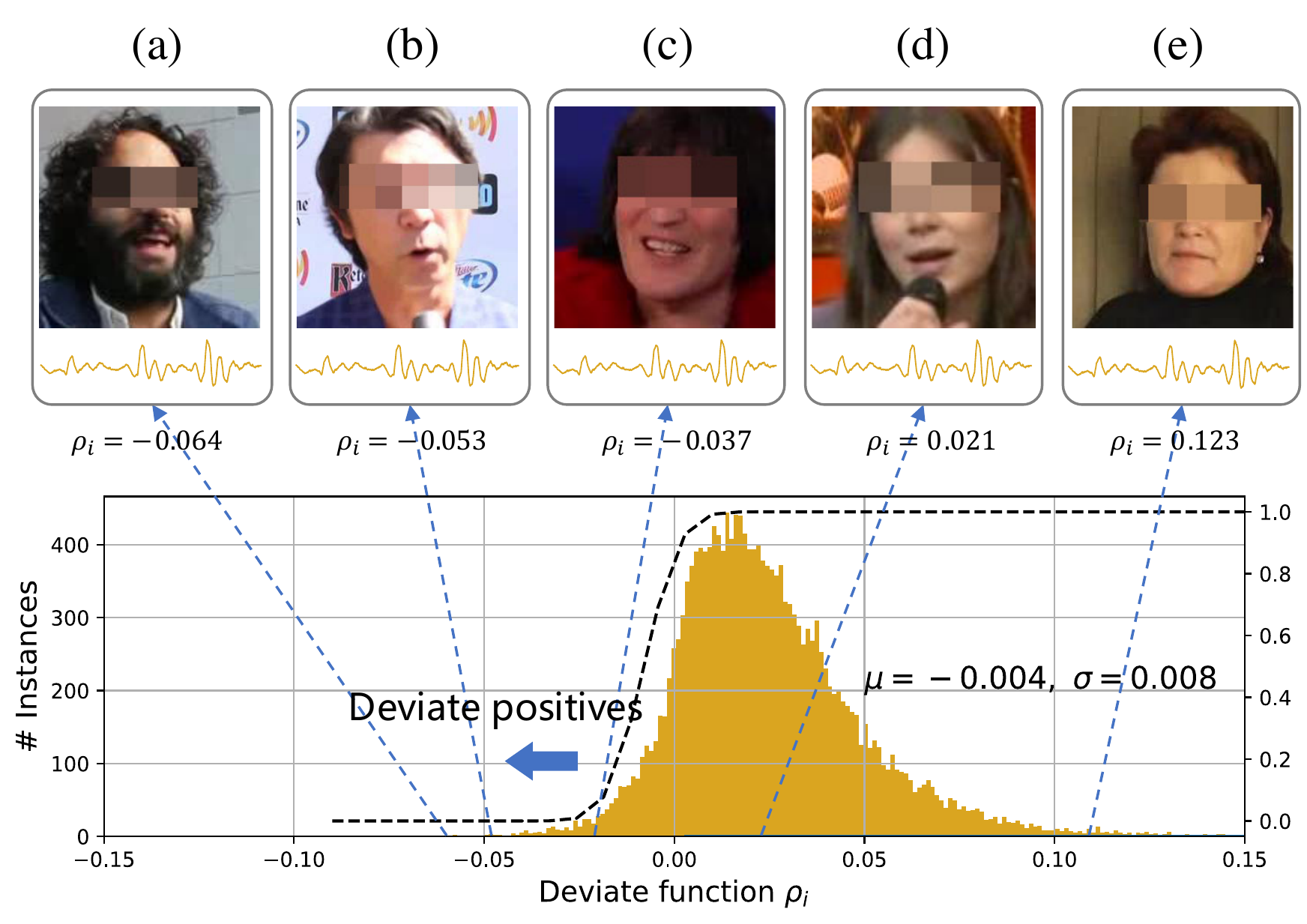} 
\caption{Deviate positives in a CMPC pretrained model. The relationship between instance numbers and $\rho(i)$ is shown in the histogram. We show some positive pairs under different $\rho(i)$ and describe the voice track: (a) Female voice in the background, (b) Noisy background with music, (c) Mature male voice, (d) Female voice, (e) Female voice. Deviate positives prone to hold lower $\rho(i)$.}
\label{fig:pos}
\end{figure}

\begin{table*}[htp]
\setlength{\tabcolsep}{6pt}
\setlength\extrarowheight{2.5pt}
\centering
\begin{tabular}{l|c|cccc|cccc|c}
\toprule[1pt]
\multirow{2}[1]{*}{Method} & \multirow{2}[1]{*}{Kind of Sup.} & \multicolumn{4}{c|}{V-F Matching (Acc \%)} & \multicolumn{4}{c|}{F-V Matching (Acc \%)} & \multirow{2}[1]{*}{Mean} \\
      &       & U     & G     & N     & GN    & U     & G     & N     & GN    &  \\
\hline
SVHF \cite{nagrani2018svhf} & Supervised & 81.0  & 63.9  & -     & -     & 79.5  & 63.4  & -     & -     & - \\

DIMNet-I \cite{wen2018dimnet} & Supervised & 83.5  & 70.9  & 82.0  & 69.9  & 83.5  & 71.8  & 82.4  & 70.9  & 76.9 \\

DIMNet-G \cite{wen2018dimnet} & Supervised & 72.9  & 50.3  & 71.9  & 50.2  & 72.5  & 50.5  & 72.2  & 50.6  & 61.4 \\

Horiguchi's \cite{horiguchi2018face} & Supervised & 78.1  & 61.7  & -     & -     & 77.8  & 60.8  & -     & -     & - \\

Wen's \textdagger \cite{wen2021seeking} & Supervised & 81.4  & 67.8  & 79.6  & 66.7  & 79.9  & 66.1  & 79.3  & 66.9  & 73.5 \\
\hline

Kim's \cite{kim2018learning} & Self-supervised & 78.2  & 62.9  & 76.4  & 61.6  & 78.6  & 61.6  & 76.7  & 61.2  & 69.7 \\

CID \cite{morgado2021agreement}   & Self-supervised & 78.3  & 64.7  & 77.8  & 64.2  & 77.6  & 64.7  & 77.2  & 64.1  & 71.1 \\

CMPC (Ours)  & Self-supervised & \textbf{82.2}  & \textbf{67.4}  & \textbf{81.1}  & \textbf{67.1}  & \textbf{81.7}  & \textbf{67.4}  & \textbf{81.1}  & \textbf{66.8}  & \textbf{74.4} \\
\hline

CID \cite{morgado2021agreement}  & Low-shot & 82.9  & 68.6  & 82.6  & 66.3  & 79.9  & 68.7  & 82.1  & 68.7  & 75.0 \\
CMPC (Ours)   &  Low-shot & \textbf{85.3}  & \textbf{70.6}  & \textbf{84.3}  & \textbf{70.2} & \textbf{84.9}  & \textbf{70.6}  & 
\textbf{84.4}  & \textbf{71.1}  & \textbf{77.7} \\
\bottomrule[1pt]
\end{tabular}%
\caption{\textbf{Matching Results}. Bidirectional (V-F and F-V) matching performance in accuracy. The U/G/N/GN represent different demographic attributes to construct testing groups. All the test data are from the unseen and unheard identities. \textdagger: It uses different face data from previous and ours, we reproduce it on VoxCeleb dataset.}
\label{tb:matching}
\end{table*}%

\subsection{Instance Recalibration}
\label{sub:ir}

To address the deviate positive problem, we propose a \textit{Instance Recalibration} mechanism to reduce the importance of deviate positives on representation learning.

For a video instance, it is more likely to be deviate if 
(1) the similarity of its voice representation $\mathbf{v}_i$ and face representation $\mathbf{f}_i$ is small, and
(2) the similarity of its corresponding voice prototype $\mathbf{c}_{s_1}^V$ and face prototype $\mathbf{c}_{s_2}^F$ is large.
Base on the above observations, we design a deviation function $\rho(\cdot)$ which measures the possibility of an instance being deviate:
\begin{equation}
   \rho(i) = \mathbf{v}_i^T\mathbf{f}_i  - \big( \frac{1}{R}\sum_{r=1}^{R}{(\mathbf{c}_{s_1}^V)^T \mathbf{c}_{s_2}^F} \big)
\end{equation}
The deviation function describes how far the instance representation similarity deviates from prototype similarity. The smaller $\rho(i)$ is, the more likely the $i$-th instance is a deviate positive. We then compute a recalibration factor $w_i$ by:
\begin{equation}
w_i = F_{\mathcal{N}}(\rho(i)).
\end{equation}
Inspired by \cite{morgado2021robust}, we use the cumulative distribution function (CDF) to control the shape of recalibration function. $F_\mathcal{N}$ is the CDF of a Gaussian distribution $\mathcal{N}(\mu+\delta\sigma, \kappa\sigma^2)$, where $\mu$ and $\sigma$ are the mean and the variance of the deviation function $\rho(i)$ on all instances, and $\delta$ and $\kappa$ are hyper-parameters to control the shape of the recalibration function. 
The recalibration factor $w_i$ is then use to rescale the CMPC loss of each instance.
The final loss function are defined as:
\begin{equation}
\mathcal{L} = \frac{1}{\sum_n{w_n}}\sum_{i}{w_i}\mathcal{L}_{\mathrm{CMPC}}(\mathbf{v}_i, \mathbf{f}_i)
\end{equation}


\section{Experiments}
\label{sec:experiments}

\subsection{Dataset}
We use the VoxCeleb dataset \cite{nagrani2017voxceleb} in our experiments, which contains 21,063 video clips of 1251 celebrities, extracted from videos uploaded to YouTube. The dataset is gender-balanced and speakers span a range of different ethnicity, accent, profession, and age. Following the previous works \cite{nagrani2018learnable,wen2021seeking}, random 1001 identities are randomly sampled as training data and the rest 250 identities as the test set. The face images are generated by facial bounding boxes cropping the video frames. The voice clips are extracted from the soundtrack of the videos with a 16kHz sampling rate at 16-bit PCM.

The test set is annotated with different demographic attributes to construct 5 testing groups, including the unstratified group (U), stratified by gender (G), stratified by nationality (N), stratified by age (A), stratified by gender and nationality (GN), and stratified by (G,N,A). The demographic groups allow us to control the important attributes in the experiments. For example, in the G-group, all the alternative instances have the same gender. The matching and verification tasks are conducted under different demographic partitions. All the evaluation lists are released.

\begin{table*}[htp]
\setlength{\tabcolsep}{6pt}
\setlength\extrarowheight{2.5pt}
\centering
\begin{tabular}{l|c|ccccc|cc}
\toprule[1pt]
\multirow{2}[1]{*}{Method} & \multirow{2}[1]{*}{Kind of Sup.} & \multicolumn{5}{c|}{Verfication (AUC)} & \multicolumn{2}{c}{Retrieval (mAP \%)} \\
   &    & U  & G  & N  & A  & GNA & V-F & F-V \\
\hline
DIMNet-I \cite{wen2018dimnet} & Supervised & 82.5 & 71.0 & 81.1 & 77.7 & 62.8 & 4.25 & 4.17 \\
DIMNet-IG \cite{wen2018dimnet} & Supervised & 83.2 & 71.2 & 81.9 & 78.0 & 62.8 & 4.42 & 4.23 \\
Horiguchi's \cite{horiguchi2018face} & Supervised & -  & -  & -  & -  & -  & 2.18 & 1.96 \\
Wen's \cite{wen2021seeking} & Supervised & 80.3 & 67.6 & 79.3 & 76.2 & 62.3 & 4.31 & 3.91 \\
\hline
PINs \cite{nagrani2018learnable} & Self-supervised & 78.5 & 61.1 & 77.2 & 74.9 & 58.8 & -  & - \\
CID \cite{morgado2021agreement} & Self-supervised & 78.2 & 65.2 & 78.4 & 75.1 & 62.9 & 3.69 & 3.55 \\
CMPC (Ours) & Self-supervised & \textbf{82.6} & \textbf{68.6} & \textbf{81.3} & \textbf{77.8} & \textbf{63.7} & \textbf{3.80} & \textbf{3.70} \\
\hline
CID \cite{morgado2021agreement}  & Low-shot & 80.8 & 67.4 & 80.7 & 77.7 & 65.3 & 4.17 & 4.10 \\
CMPC (Ours)   & Low-shot & \textbf{84.6} & \textbf{71.6} & \textbf{82.5} & \textbf{79.8} & \textbf{65.1} & \textbf{4.31} & \textbf{4.27} \\
\bottomrule[1pt]
\end{tabular}%
\caption{\textbf{Verification and Retrieval Results}.}
\label{tb:verif}
\end{table*}%

\begin{table}[htbp]
\setlength{\tabcolsep}{6pt}
\setlength\extrarowheight{2.5pt}
\centering
\begin{tabular}{ccccc}
\toprule[1pt]
\multirow{2}{*}{\begin{tabular}[c]{@{}c@{}}Cross-modal\\ Prototype\end{tabular}} &
  \multirow{2}{*}{\begin{tabular}[c]{@{}c@{}}Instance\\ Recalibration\end{tabular}} &
  \multicolumn{2}{c}{Matching} &
  \multirow{2}{*}{Verification} \\ \cmidrule{3-4}
  &   & V-F  & F-V  &      \\
\hline
\ding{55} & \ding{55} & 78.3 & 77.6 & 78.2 \\
\ding{52} & \ding{55} & 81.9 & 80.1 & 80.4 \\
\ding{52} & \ding{52} & 82.2 & 81.7 & 82.6 \\
\bottomrule[1pt]
\end{tabular}
\caption{\textbf{Ablation study}. We quantify the individual contributions of our proposed approach: cross-modal prototype and instance recalibration.}
\label{tb:ablation}
\end{table}

\begin{figure*}[t]
\centering
\includegraphics[width=2.0\columnwidth]{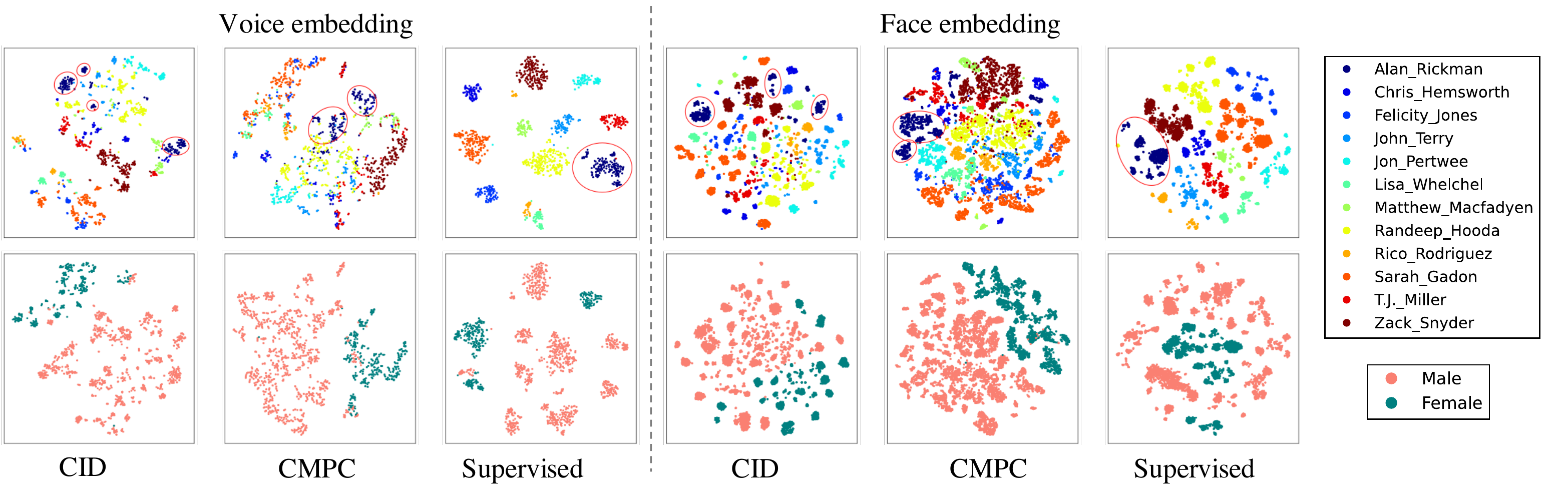}
\caption{T-SNE visualization of voice and face representations. We compare our approach with CID and the fully supervised approach. We randomly choose 12 identities' videos. Points in the figure correspond to a face frame or voice clip representation. Best view in color.}
\label{fig:tsne}
\end{figure*}

\section{Evaluation Protocols}
To appraise our learned representations, following protocols are conducted for evaluation:

\paragraph{Matching.} We are given a probe from one modality, and two alternatives from another modality, including one that belongs to the same person as the probe, and another imposter that comes from another person. The task is to affirm which alternative matches the probe. We report the bidirectional matching performance, namely voice to face (V-F) matching accuracy and face to voice (F-V) matching accuracy.

\paragraph{Verification.} We are given a voice clip and face image. The task is to judge whether they belong to the same identity. The basis of judgment is to compare the similarity of two input features with a given threshold. With different thresholds, we could get various false positive rates and true positive rates. The final performance is reported by the \textit{Area Under the ROC curve} (AUC).

\paragraph{Retrieval.} Given a probe from one modality, and a large size gallery which contains candidates from another modality. One or more candidates might match the probe. The task is to order the gallery such that the matched candidates lie at the top of the ordering. We report the bidirectional performance in terms of \textit{Mean Average Precision} (mAP).

\subsection{Implementation Details}

\paragraph{Voice and face preprocessing.}
We randomly crop a 5 seconds length speech and extract the commonly used \textit{logmel} feature which is the mel-spectrogram at log-scale during training.
The \textit{logmel} features are transformed with 100ms windows width, 10ms frame shift and 64 mel-bins, yielding an input at size of $64 \times 500$. No further speech-specific preprocessing (e.g. noise suppression, voice activity detection) is applied.
The face frames are extracted at 1fps from the video stream and cropped surrounding a human face. The input is a series of RGB images resized at $224 \times 224$ pixels. The only augmentation used on the face frames is a horizontal flip.

\paragraph{Network architecture.}
The ResNet-34 \cite{ResNet} is adopted as the voice and face encoder network. Our network is trained from scratch without any pretrained data. The original fully-connected layer of each encoder is removed. The weights of the two encoders are not shared, while they are trained jointly by the approach introduced above. We get the features after the last convolutional layer as the representations, and they are average pooled to the 512-dimensions. For both the voice and face representations, they are projected normalized into the unit sphere by L2 norm. In the evaluation phase, the projection network is removed, and only calculates the similarity between learned representations.

\paragraph{Optimization.}
We use the Adam optimizer, with a momentum of 0.9, a weight decay of 0.002 and a batch size of 128. We train for 3.2M iterations. We use a cosine annealing schedule with warm-up to tune our learning rate. The initial learning rate is 1e-4 and increases linearly to 5e-3 in the first 0.3M iterations, then gradually decreases to 1e-4. We also warmup our network in these 0.3M iterations using CID loss. The momentum coefficient in computing instance feature memory is 0.5.
We set temperature $\tau$ to 0.03, number of clusters $K=\{500, 1000, 1500\}$. The $\sigma$ and $\kappa$ in the recalibration are set to -1 and 0.1 respectively. We use KeOps \cite{KeOps} for the efficient \textit{k}-means clustering.

Our code, model and testing resources are released\footnote{\url{https://github.com/Cocoxili/CMPC}}.

\subsection{Quantitative Results}


We now compare our CMPC to prior works, including supervised and unsupervised methods on various voice-face association evaluation protocols. we can outperform previous state-of-the-art unsupervised methods on various voice-face association evaluation protocols.

Table \ref{tb:matching} summarizes the cross-modal matching performance, and Table \ref{tb:verif} presents the cross-modal verification and retrieval results.
We could see that our CMPC outperforms the previous unsupervised approaches by a large margin, bridging the gap between unsupervised and supervised methods. As this task is challenging, even human performance is around 80\% \cite{nagrani2018svhf} on the matching protocol. Our results indicate that it is feasible to learn robust voice-face representation through unlabeled videos.

\paragraph{Compared methods.}
Among the compared methods, DIMNet \cite{wen2018dimnet} uses one or more covariates, such as gender and identity, by a multi-task object function. Due to the use of more supervised information, it has achieved good results. Wen's \cite{wen2021seeking} adaptively re-weights identity samples, while its re-weighting method is based on the identity label which is not used in our method. Meanwhile, their face data are obtained from the VGGFace dataset, which is different from the previous and our method. The results we report are reproducing \footnote{https://github.com/KID-7391/seeking-the-shape-of-sound} their method on the VoxCeleb dataset. PINs \cite{nagrani2018learnable} also employs a contrastive loss to learn the representation under an unsupervised setting, while our contrastive loss is distinct from theirs in which a positive pair only compares with one negative pair at once. It can be seen as a variant of the triplet loss.

\paragraph{Low-shot learning.}
To further evaluate the learned representations, we conduct our experiments on the low-shot training, which only uses a few instances per-identity. These instances are labeled by the identity information. We follow the DIMNet \cite{wen2018dimnet} setup to finetune our network, while we do not use other covariates such as gender, nationality, etc. We representatively select 5 instances per-identity, which is about one-third of the whole dataset. We report the evaluation results in Table \ref{tb:matching} and Table \ref{tb:verif}. Due to the better voice and face representations we learned, the results even outperform those with entire dataset supervision.

\paragraph{Ablation Study.}
To shed light on the contributions of our two components, cross-modal prototype contrast and instance recalibration, we report the matching and verification results on the unseen-unheard unstratified group (U) of VoxCeleb dataset as in Table \ref{tb:ablation}. The other experimental settings are kept identical. Both two components yield a positive impact on the results, individually and jointly.
\subsection{Qualitative Analyses}

In Figure \ref{fig:tsne}, we visualize the learned representations of voice clips and face images using t-SNE. We color the sample points by identity labels and gender labels respectively. We compare our approach with the CID method \cite{morgado2021agreement} and fully supervised approach \cite{wen2018dimnet}.

The CID only focuses on the local association between instances and ignores the semantic structure. It undesirable pushes false-negative instances apart. Therefore, different instances representations of the same person will appear far away, such as the samples marked in the red circles. While, in the supervised method, every identity's representations are brought together tightly. In our method, we keep both voice and face representations of the same identity closer together even without using any labels, due to the cross-modal prototype contrast.
Another interesting thing, the representations are not well separated by gender even in the supervised method, as we do not use gender supervision. While in our approaches, they are separated naturally.



\section{Conclusion}
\label{sec:conclusion}

In this paper, we propose a novel unsupervised approach to learn voice-face association from the talking-face videos. We use the natural relevance between audio and visual frames to establish a self-supervised approach. Our method is based on the framework of contrastive learning, where the selection of positive and negative samples is crucial.
We reveal their adverse impact by analyzing their gradient flow in the training phase. To tackle the false negative problem, we propose a cross-modal prototype contrast algorithm by estimating the semantic prototypes of each instance, then we replace the instance-to-instance contrast with instance-to-prototype contrast in a cross-modal way. To tackle the deviate positive problem, we dynamically recalibrate the instances by a deviation function. Experiments on cross-modal matching, verification, and retrieval have shown the superiority of our learn representations.

\section*{Acknowledgments}

This work is supported by the major Science and Technology Innovation 2030 ``New Generation Artificial Intelligence'' project 2020AAA0104803.




\bibliographystyle{named}
\bibliography{refs_ijcai22}

\end{document}